\documentclass[seceqn]{elsart}

\usepackage{amssymb}
\usepackage{graphicx}
\usepackage{epsfig}

\begin{document}


\title{Parallelization of Markov Chain Generation 
and its Application to the Multicanonical Method}

\author[sugihara]{Takanori Sugihara},
\author[higo]{Junichi Higo}, and
\author[nakamura]{Haruki Nakamura}

\address[sugihara]{Next-Generation Supercomputer R\&D Center, RIKEN, \\
2-1-1 Marunouchi, Chiyoda, Tokyo 100-0005, Japan}
\address[higo]{MEI Center, Osaka University, \\
2-2 Yamadaoka, Suita, Osaka 565-0871, Japan}
\address[nakamura]{Institute for Protein Research, Osaka University, \\
3-2 Yamadaoka, Suita, Osaka 565-0871, Japan}

\begin{abstract}
We develop a simple algorithm to parallelize 
generation processes of Markov chains. 
In this algorithm, multiple Markov chains are generated in parallel 
and jointed together to make a longer Markov chain. 
The joints between the constituent Markov chains are processed 
using the detailed balance. 
We apply the parallelization algorithm to multicanonical calculations 
of the two-dimensional Ising model and demonstrate accurate 
estimation of multicanonical weights. 
\end{abstract}


\section{Introduction}
\label{intro}

The MCMC (Markov Chain Monte-Carlo) method \cite{metropolis} 
has played an important role in study of complex systems 
with many degrees of freedom. 
For example, MCMC has been applied to various many-body problems 
such as proteins \cite{protein}, 
spin systems \cite{suzuki}, and lattice gauge theory \cite{creutz}. 
Although the method has achieved great success, 
there are systems where Monte-Carlo sampling does not work 
due to local minima of energy functions.
For example, in analysis of protein folding, Monte-Carlo 
random walks are trapped in narrow regions of energy space 
because there are so many local minima. 
For large proteins, it is a hard problem to obtain sufficiently 
large number of statistical samples because it takes very long time 
for a complicated conformation to escape from a local minimum 
in energy space.

The multicanonical method \cite{berg1,berg2,iba} 
may be useful for resolving the local 
minimum problems. It is a kind of generalized-ensemble methods. 
It has been applied to the above mentioned problems and worked nicely. 
Especially its application to protein folding with the multicanonical 
MD \cite{hoe,nakajima} has been remarkable. 
Recently, Ikebe et al have succeeded in folding calculation 
of a 40-residue protein \cite{ikebe}. 
Kamiya et al have demonstrated flexible molecular docking 
between a protein and a ligand \cite{kamiya}. 
Since protein structure prediction is one of the most important 
subjects in life sciences, it has been expected that the multicanonical 
method will make a large progress in clarification of the fundamental 
laws of life and development of high-performance drugs. 

The multicanonical method is based on an artificial ensemble 
that gives a flat probability distribution in energy space. 
The advantage of the method is enhancement of rare configurations, 
which results in random walks in wide range of energy space. 
In this method, multicanonical weights are estimated 
in an iterative way. To estimate a multicanonical weight, 
one has to generate a Markov chain many times, which must be 
sufficiently long for accurate estimation. 
The number of arithmetic operations increases exponentially 
as the number of amino-acid residues becomes large. 
Since actual proteins are composed of more than 
a hundred amino-acid residues, 
the number of necessary operations for folding simulation is huge. 
It would be a possibility to decrease the execution time 
using massively parallel supercomputers. 

In this paper, we propose an algorithm to generate a Markov chain 
in a parallel way. In MCMC calculations, the detailed balance is 
checked between the last and a newly generated configuration 
to determine acceptance or rejection of the new one. 
Accepted configurations constitute a so-called Markov chain. 
A Markov chain is a one-dimensional object that is generated 
serially. Since acceptance or rejection of a new configuration 
depends on the last configuration, naive algorithm for Markov-chain 
generation such as the Metropolis method is essentially serial. 
At a glance, it seems that 
the MCMC algorithm cannot be parallelized. 
In this paper, we are going to show a method to joint multiple 
Markov chains together to make a longer Markov chain. 
The constituent Markov chains are independently generated 
starting from different initial configurations. 
The essential point of the method is that joints between chains 
are processed so that the detailed balance is satisfied. 
Based on the detailed balance, unnecessary configurations are 
discarded from each Markov chain. 
The remaining Markov chains are connected 
together to make a longer Markov chain. 
By repeating this procedure, one can increase the length 
of the Markov chain arbitrarily. 
In this algorithm, other operations such as evaluation of 
energy functions are not parallelized. 

To demonstrate how the parallelization algorithm works, we solve 
the two-dimensional Ising model by combining the proposed 
parallelization algorithm and the multicanonical method. 
Multicanonical weights are estimated from histograms, 
which are obtained using the parallelization algorithm. 

In Sec. \ref{para}, we introduce an algorithm to generate 
a long Markov chain in a parallel way. 
In Sec. \ref{multic}, we review  the multicanonical method briefly. 
In Sec. \ref{app}, we apply the parallelization algorithm to 
the two-dimensional Ising model with the multicanonical method. 
In Sec. \ref{results}, we show numerical results. 
Sec. \ref{conclusions} is devoted to conclusions.

\section{Parallelization of Markov-Chain Generation}
\label{para}
Let us consider a Markov chain that is composed of 
$M$ configurations, 
\begin{equation}
C_1,C_2, \dots, C_M. 
\end{equation}
We call $C_k$ a configuration, which is a set of values of multiple 
variables. $C_k$ are generated so that configurations distribute 
according to a specified probability distribution $P(C)$. 
In actual calculations, the number of configurations is finite. 
Therefore, a set of configurations only reproduces $P(C)$ approximately. 
The associated errors vanishes in the limit $M\to \infty$. 

In MCMC, configurations are generated serially so that 
the detailed balance is satisfied. One determines acceptance 
or rejection of a randomly generated configuration $C_k$ by 
checking the detailed balance between $C_{k-1}$ and $C_k$. 
At a glance, the MCMC algorithm seems to be essentially serial and 
is not suited to parallel computation. In order to decrease 
execution time using parallel computers, we propose 
a simple algorithm to parallelize Markov-chain generation. 

Let us consider a computer of which parallelism is $p$. 
In our parallelization algorithm, each computing node generates 
a Markov-chain separately. 
In order to obtain a much longer Markov chain having $M_{\rm total}$ 
configurations than a Markov chain having $M_{\rm node}$ 
configurations generated by each computing node, the $p$ Markov chains 
are connected satisfying the detailed balance condition as follows:

\vspace{0.3cm}
\noindent
{\it Algorithm}
\begin{description}
\item[1.] Set the total histogram zero, $h(E)=0$, 
where $E$ is energy variable. 
Prepare initial configurations $C_1^{(i)}$ randomly 
on $i$-th node ($i=0,\dots,p-1$).

\item[2.] In each node, generate a Markov chain 
composed of $M_{\rm node}$ configurations, 
\begin{equation}
C_1^{(i)}\to C_2^{(i)}\to \cdots C_{M_{\rm node}}^{(i)}, 
\quad i=0,\dots,p-1. 
\end{equation}

\item[3.] Set $i=0$.

\item[4.] In the $i$-th node, check the detailed balance between 
$C_{M_{\rm node}}^{(n)}$ and $C_k^{(i)}$, 
where $C_{M_{\rm node}}^{(n)}$ with $n=(i-1+p)$ mod $p$ 
is the last configuration of the previous node. 
The index $k$ is increased from $1$ to $M_{\rm node}$ 
till the detailed balance is satisfied. 
When there is a configuration that satisfies the detailed balance, 
that configuration and the succeeding ones are accepted. 
When there is no accepted configuration, 
set $C_{M_{\rm node}}^{(i)} \leftarrow C_{M_{\rm node}}^{(n)}$. 
Make a local histogram $h_{\rm local}(E)$ 
using only the accepted configurations. 
Calculate the total histogram, 
$h(E) \leftarrow h(E)+h_{\rm local}(E)$. 
Send the values of $h(E)$ and the energy of 
$C_{M_{\rm node}}^{(i)}$ to the 
$((i+1)\; {\rm mod}\; p)$-th node. 
(When $i=0$, $C_{M_{\rm node}}^{(p-1)}$ of the previous 
iteration is used for checking of the detailed balance.)

\item[5.] Set $i\leftarrow i+1$. If $i<p$, go to step 4. 

\item[6.] If the total number of the configurations contained 
in the total histogram $h(E)$ is smaller than the 
specified value $M_{\rm total}$, adjust $M_{\rm node}$ 
appropriately so that additional calculations are minimized, 
and then go to step 2. 
\end{description}

We are going to give some remarks on the algorithm below. 

The algorithm produces a Markov chain of which the configuration 
number is $M_{\rm total}$. Actually, the algorithm outputs 
the total histogram $h(E)$. 
As shown in the next section, the obtained histogram $h(E)$ 
is used to estimate a multicanonical weight $w(E)$. 
The algorithm is repeated certain times till 
a multicanonical weight that covers a sufficiently large 
energy area is obtained. 
Since the algorithm only assumes the detailed balance and 
does not depend on the details of probability distribution, 
the algorithm works in any MCMC calculations. 

In each node, the last configuration generated in the previous 
iteration is used for generating a Markov chain 
as an initial configuration in the next iteration 
including when the considered weight has been updated.
This means that $p$ Markov chains are generated 
independently from first to last. 
Equilibrium depends on the current weight.

There may be no configuration that satisfies the detailed 
balance in step 4. In this case, all of the configurations 
contained in that computing node is discarded, 
which results in low efficiency of Markov-chain generation. 
One can increase acceptance rate by ordering 
computing nodes according to energy values 
of generated configurations. 
We will pursue this technique in the next paper.

\section{The Multicanonical Method}
\label{multic}

\subsection{Estimation of Multicanonical Weights}
\label{weight}

We are going to combine the algorithm introduced in the previous 
section with the multicanonical method. 
Let us review the multicanonical method briefly. For the details 
of the multicanonical method, see \cite{berg1,berg2,iba}.

Consider a statistical system that is defined with 
a Boltzmann weight
\begin{equation}
w_{\rm B}(E) = e^{-\beta E},
\end{equation}
where $\beta=1/k_{\rm B}T$. 
Hereafter, we set $k_{\rm B}=1$ for simplicity. 

Probability distribution is given by
\begin{equation}
P(E)=c_\beta D(E) w_{\rm B}(E),
\end{equation}
where $D(E)$ is density of states. 
The constant $c_\beta$ is determined with 
the normalization condition$\sum_E P(E)=1$. 

In the multicanonical method, an extended weight $w_{\rm M}(E)$ is 
defined so that multicanonical distribution is independent of energy, 
\begin{equation}
P_{\rm M}(E)=c_{\rm M} D(E) w_{\rm M}(E) \sim c_{\rm M}.
\end{equation}

We can obtain the multicanonical weight $w_{\rm M}$ 
in a recursive way. 
We denote the $n$-th multicanonical weight as $w^{(n)}$, 
where $n=1,2,\dots,I$. 
In the initial step, we set $w^{(1)}=1$, which corresponds to 
high-temperature limit of the Boltzmann weight $w_{\rm B}$. 
With the $w^{(n)}$, we generate a sufficiently long Markov chain and 
obtain a total histogram $h^{(n)}(E)$. 
Then, we generate the next weight $w^{(n+1)}$ 
using the current weight $w^{(n)}$ and 
the obtained histogram $h^{(n)}(E)$. 
\begin{equation}
w^{(n+1)}(E) =
\cases{
w^{(n)}(E), & if $h^{(n)}(E) = 0$, \cr
\displaystyle\frac{w^{(n)}(E)}{h^{(n)}(E)}, & otherwise. \cr
\label{new_weight}
}
\end{equation}
We expect that the weight $w^{(n)}$ approaches to the correct 
multicanonical weight $w_{\rm M}$ 
if this process is repeated certain times. 

To obtain the multicanonical weight accurately, sufficiently 
large statistics are necessary for generating $h^{(n)}(E)$. 
We can decrease execution time consumed to generate 
sufficiently long Markov chains by making use of the 
parallelization algorithm introduced in Sec. \ref{para}.
\cite{note}.

\subsection{Evaluation of Statistical Average}
\label{stat}
We can calculate statistical average at arbitrary temperature 
with the following reweighting formula: 
\begin{equation}
\langle O(x) \rangle \equiv \sum_x O(x) P(x) =
\frac{\sum_i O(x^{(i)}) D(E(x^{(i)})) e^{-\beta E(x^{(i)})} }
{\sum_j D(E(x^{(j)})) e^{-\beta E(x^{(j)})}},
\label{reweighting}
\end{equation}
where $x$ represents multiple variables and 
$x^{(i)}$ is a configuration. 
If the operator $O(x)$ can be represented as a function of energy $E$ 
and the multicanonical weight $w_{\rm M}(E)$ is known, 
one can evaluate Eq. (\ref{reweighting}) without configurations. 
In this case, the formula (\ref{reweighting}) reduces to
\begin{equation}
\langle O(x) \rangle =
\frac{\sum_E O(E) w_{\rm M}(E)^{-1} e^{-\beta E} }
{\sum_E w_{\rm M}(E)^{-1} e^{-\beta E}},
\label{reweighting2}
\end{equation}
because the density of states is inversely proportional to the weight 
($D(E)\propto 1/w(E)$) and the histogram has flat distribution. 
In Eq. (\ref{reweighting2}), we can evaluate $\langle O(x) \rangle$ 
by taking the summation for all possible energy values $E$.
The reduced reweighing formula (\ref{reweighting2}) cannot be used 
for quantities that are dependent on local variables. 

\section{Application to the 2D Ising Model}
\label{app}
We are going to solve the two-dimensional Ising model at finite 
temperature combining the parallelization algorithm 
(Sec. \ref{para}) and the multicanonical method (Sec. \ref{multic}). 

\subsection{The 2D Ising Model}
\label{ising}
The model is defined on two dimensional square lattices. 
The number of lattice sites is $N=L^2$, where $L$ is the lattice size. 
We assume periodic boundary conditions. 
The Hamiltonian of the model is defined as follows:
\begin{equation}
H = -\sum_{\langle i,j \rangle} s_i s_j,\quad
s_i = \pm 1, 
\end{equation}
where the summation is taken for all possible 
nearest-neighbour sites. 
All information of thermodynamics is contained in 
the partition function
\begin{equation}
Z = \sum_{s_1,\dots,s_N} e^{-\beta H}. 
\end{equation}
Based on this, we calculate energy $E$, specific heat $C$, 
free energy $F$, and entropy $S$ in a statistical way.
\begin{eqnarray}
E &\equiv& \langle H \rangle, \\
C &\equiv& \frac{dE}{dT}
= \beta^2 (\langle H^2 \rangle - \langle H \rangle^2), \label{specific}\\
F &\equiv& -\frac{1}{\beta}\ln Z, \\
S &\equiv& \beta (E-F).
\end{eqnarray}

\subsection{Coding}
\label{code}
We apply a combination of the parallelization algorithm and the 
multicanonical method to the two-dimensional Ising model. 
We estimate multicanonical weights iteratively 
with the parallelization algorithm. 
We implement the codes with FORTRAN77. 
Our code set is composed of the following two parts: 
(i) generation of the multicanonical weight and 
(ii) evaluation of statistical average. 
These are implemented as separate two codes. 
Since almost all of the arithmetic operations are contained 
in the part (i), only the part (i) is parallelized 
with MPI-1, which is a library specification for message-passing 
proposed as a standard \cite{mpi}. The obtained multicanonical 
weight is used as an input to the part (ii). 
We calculate statistical average of energy, specific heat, 
free energy, and entropy. Since these quantities can be 
represented as functions of energy, 
we can use the reweighting formula (\ref{reweighting2})
in the part (ii). Execution time of the part (ii) is very short 
like a second on a Xeon 1.50 GHz processor. 

We store logarithm of the weight, not the weight itself, 
because the absolute values of the weight may be very small. 
For this reason, we perform all necessary operations under 
logarithm to evaluate statistical quantities. 

Each node generates a Markov chain composed of $M_{\rm node}$ samples 
using the current multicanonical weight. 
When all $p$ nodes have done Markov-chain generation, the obtained chains 
are linked together using the detailed balance as explained before. 
To be concrete, an energy value of the last configuration 
and the total histogram $h(E)$ 
are sent to the next node for the detailed-balance checking 
and histogram generation. 
The Metropolis method is used for the detailed-balance checking.
If the total number of accepted configurations contained in the 
total histogram $h(E)$ is smaller than the specified sample number 
$M_{\rm total}$, the chain generation process is repeated. 
If it is larger than $M_{\rm total}$, the next weight 
is calculated using Eq. (\ref{new_weight}) 
and distributed to all the nodes. 
Then, a new iteration process is started with the new weight. 
The above iteration process for weight estimation is repeated 
the specified $I$ times.

\section{Numerical Results}
\label{results}

\begin{figure}
\begin{center}
\includegraphics[width=60mm]{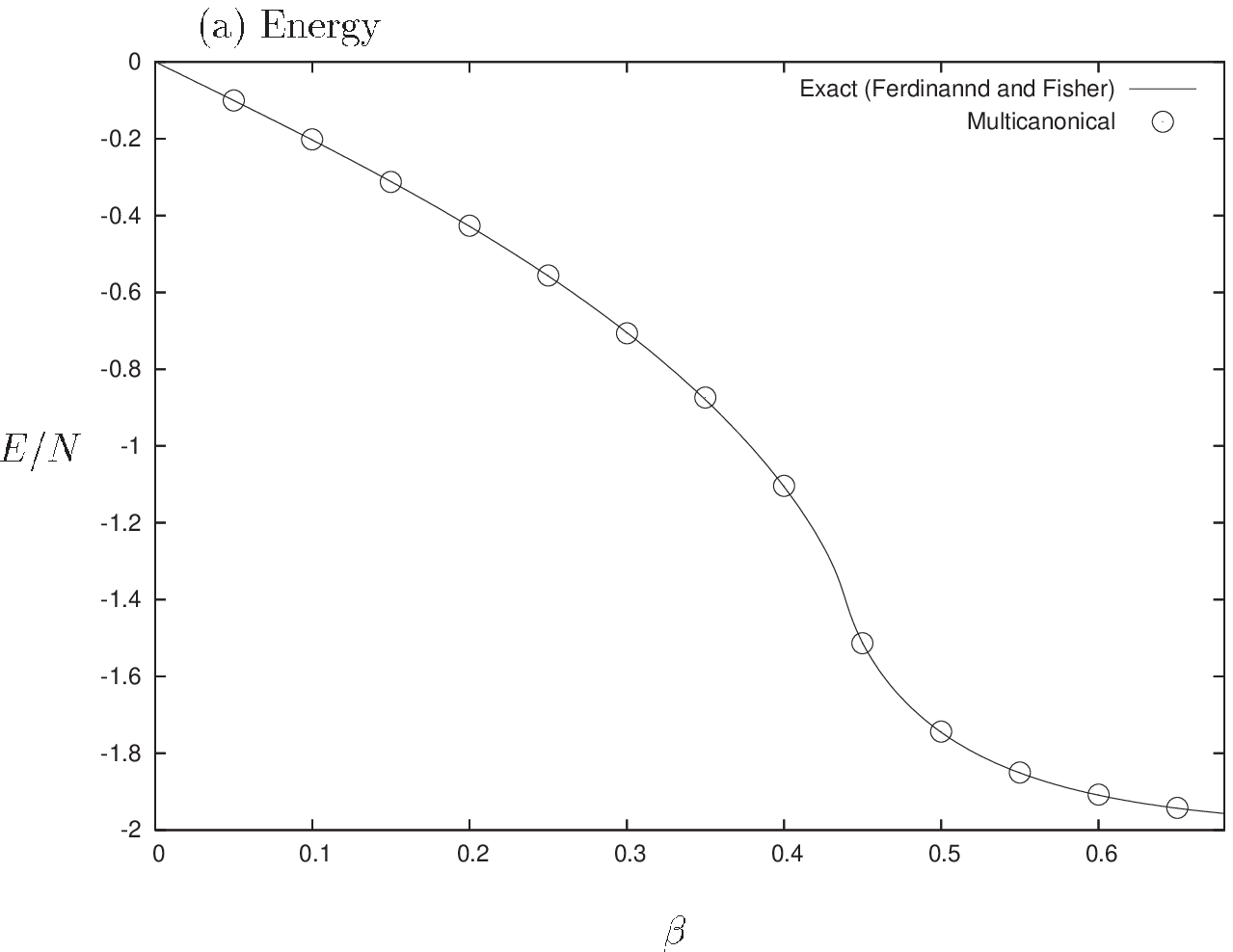}
\includegraphics[width=60mm]{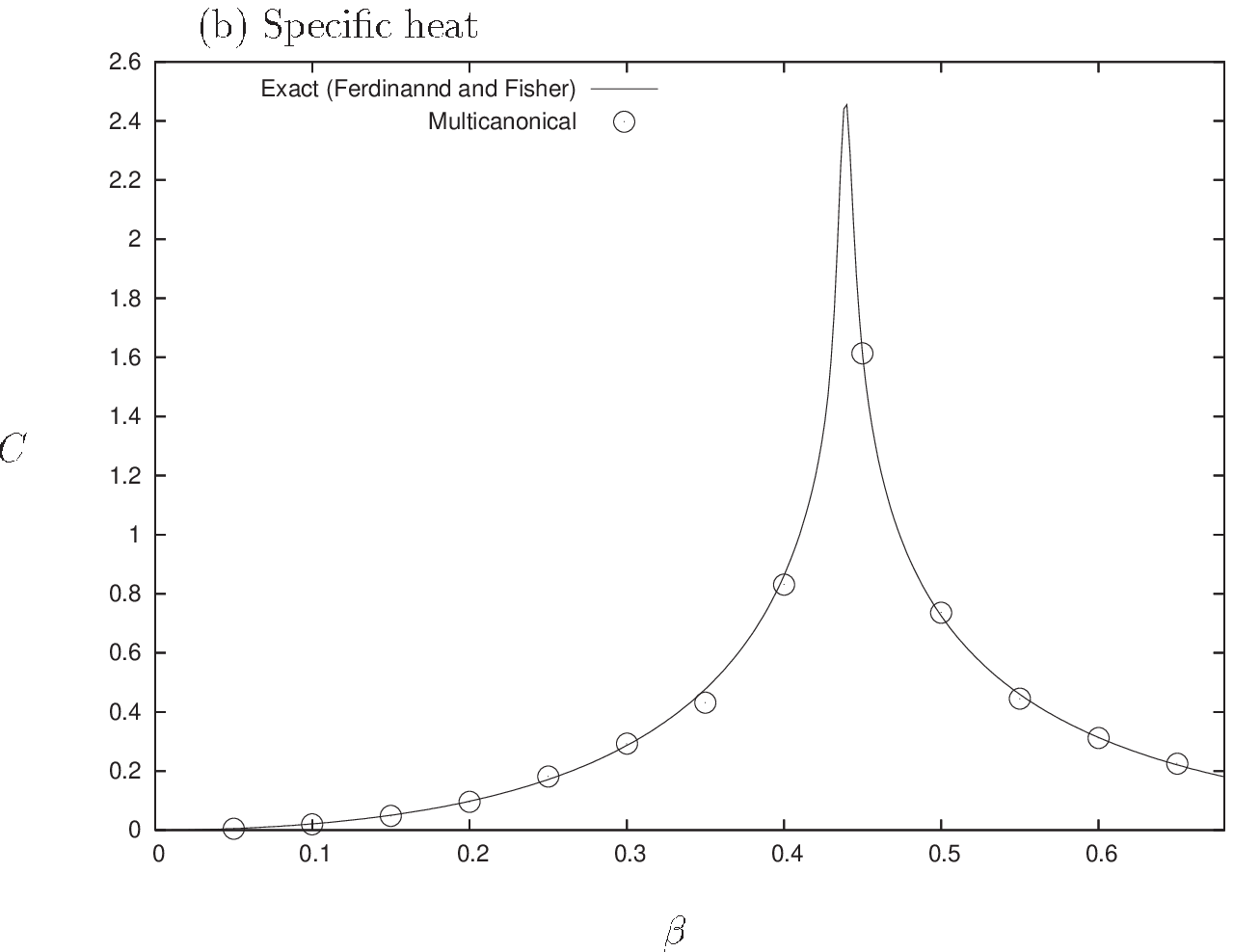}
\includegraphics[width=60mm]{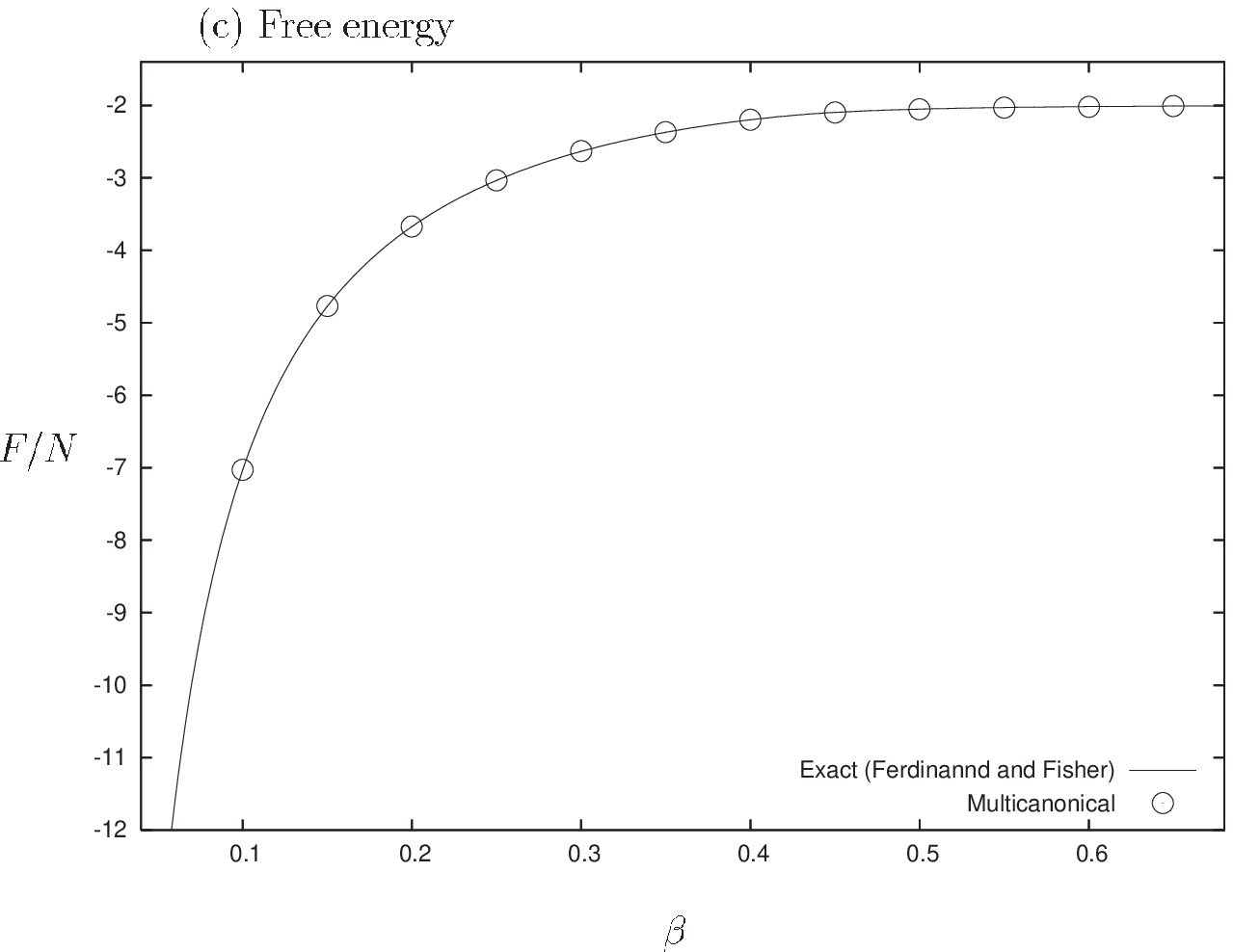}
\includegraphics[width=60mm]{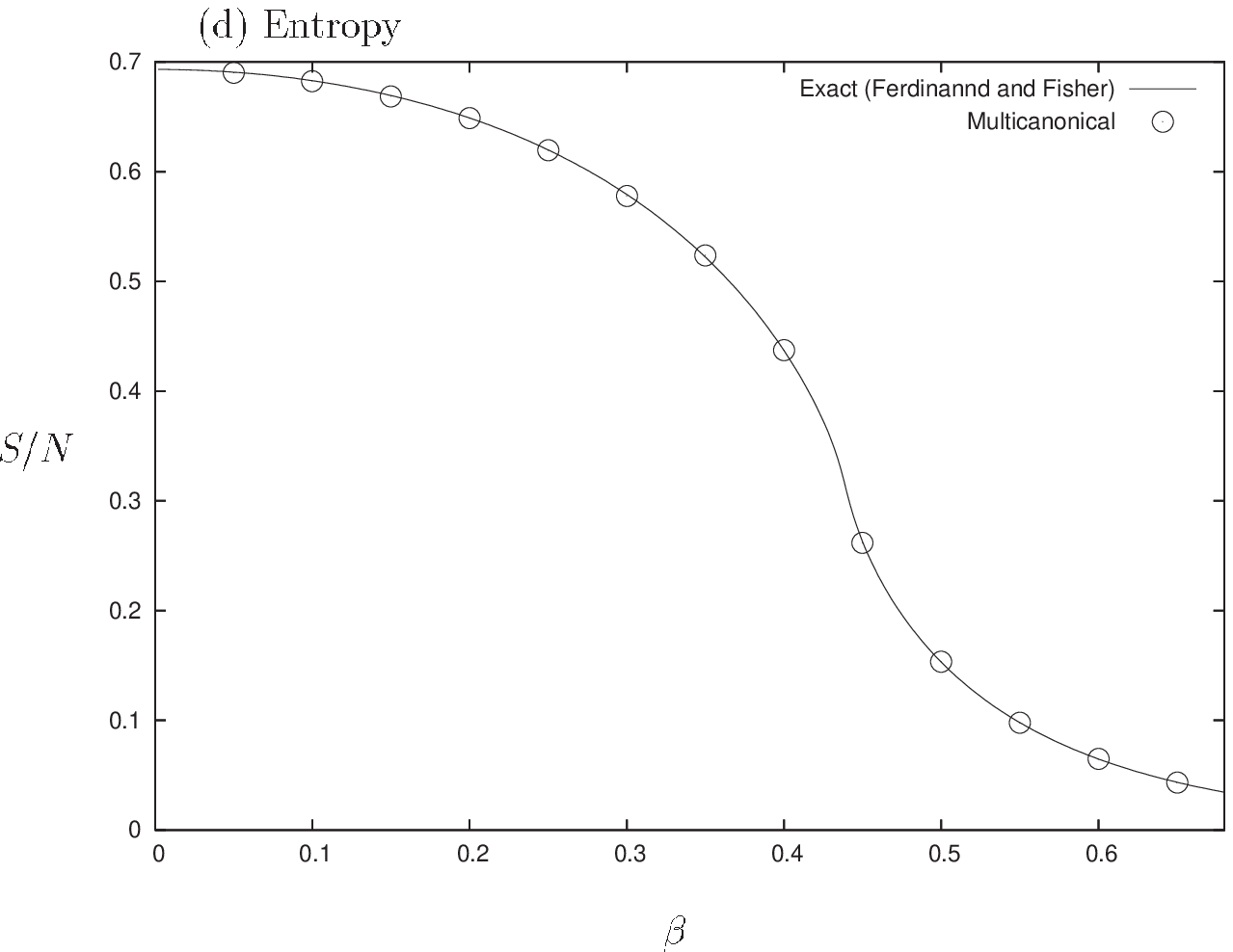}
\end{center}
\caption{For $L=100$, 
(a) energy $E/N$, (b) specific heat $C$, 
(c) free energy $F/N$, and (d) entropy $S/N$ 
are plotted as functions of inversed temperature $\beta$. 
The exact results given by Ferdinand and Fisher \cite{ferdinand}
and ours are plotted with solid lines and circles, 
respectively. 
\label{fig1}}
\end{figure}

With the implemented codes, we generate multicanonical weights 
using the parallelization algorithm and calculate 
statistical quantities using Eq. (\ref{reweighting2}).  
Table \ref{parameter} is a list of parameter values used to 
plot Fig. \ref{fig1}
\begin{table}[htb]
\vspace{0.5cm}
\caption{Parameter values used for calculation of 
statistical quantities shown in Fig. \ref{fig1}.}
\vspace{0.2cm}
\label{parameter}
\begin{tabular}{lcc}
\hline
Parameter & Value & Meaning\\
\hline
$L$ & $100$ & Lattice size \\
$N$ & $L^2$ & The number of lattice sites \\
$p$ & $32$ & Parallelism \\
$M_{\rm total}$ & $10^8$ & The total number of samples for one iteration\\
$M_{\rm node}$ & $M_{\rm total}/p$ & The number of samples generated by a node for one iteration \\
$I$ & $1300$ & The number of iterations for weight generation\\
\hline
\end{tabular}
\end{table}

In Fig. \ref{fig1}, 
we compare our results with the exact finite-lattice ones 
obtained by Ferdinand and Fisher \cite{ferdinand} 
when lattice size is $L=100$. 
As shown in figures (a), (c), and (d), our results for 
energy, free energy, and entropy agree with 
the exact ones very well. On the other hand, 
in Fig. \ref{fig1} (b), basically two results agree 
but we admit slight difference. 
In general, specific heat is more sensitive to errors associated 
with obtained multicanonical weights than energy as seen 
in the definition of specific heat (\ref{specific}).

When parallelism $p$ is small like $1\le p \le 1000$, 
the number of samples generated by each node $M_{\rm node}$ is 
sufficiently large with a fixed $M_{\rm total}$. 
In this case, errors associated with multicanonical weights 
would be small even if the detailed 
balance is not imposed on the joints of short Markov chains 
because the number of joints is much smaller than 
the total number of samples. 
However, in near future, supercomputers will acquire 
parallelism of several millions or more. 
For example, if the same code is executed with 
$p=10^6$ and $M_{\rm total}=10^8$, we have $M_{\rm node}=100$, 
which is quite small. 
In this case, there are relatively many joints 
compared to the total number of samples. 

In order to confirm that the proposed algorithm works well 
even when $M_{\rm node}$ is very small, 
i.e. each constituent Markov chain is very short, 
we perform a simple experiment. 
In this experiment, we generate very short Markov chains 
with $M_{\rm node}=100$, and just connect the chains 
together without imposing the detailed balance on the joints. 
This procedure is repeated till the specified number of 
samples have generated to make a multicanonical weight.

\begin{figure}
\begin{center}
\includegraphics[width=100mm]{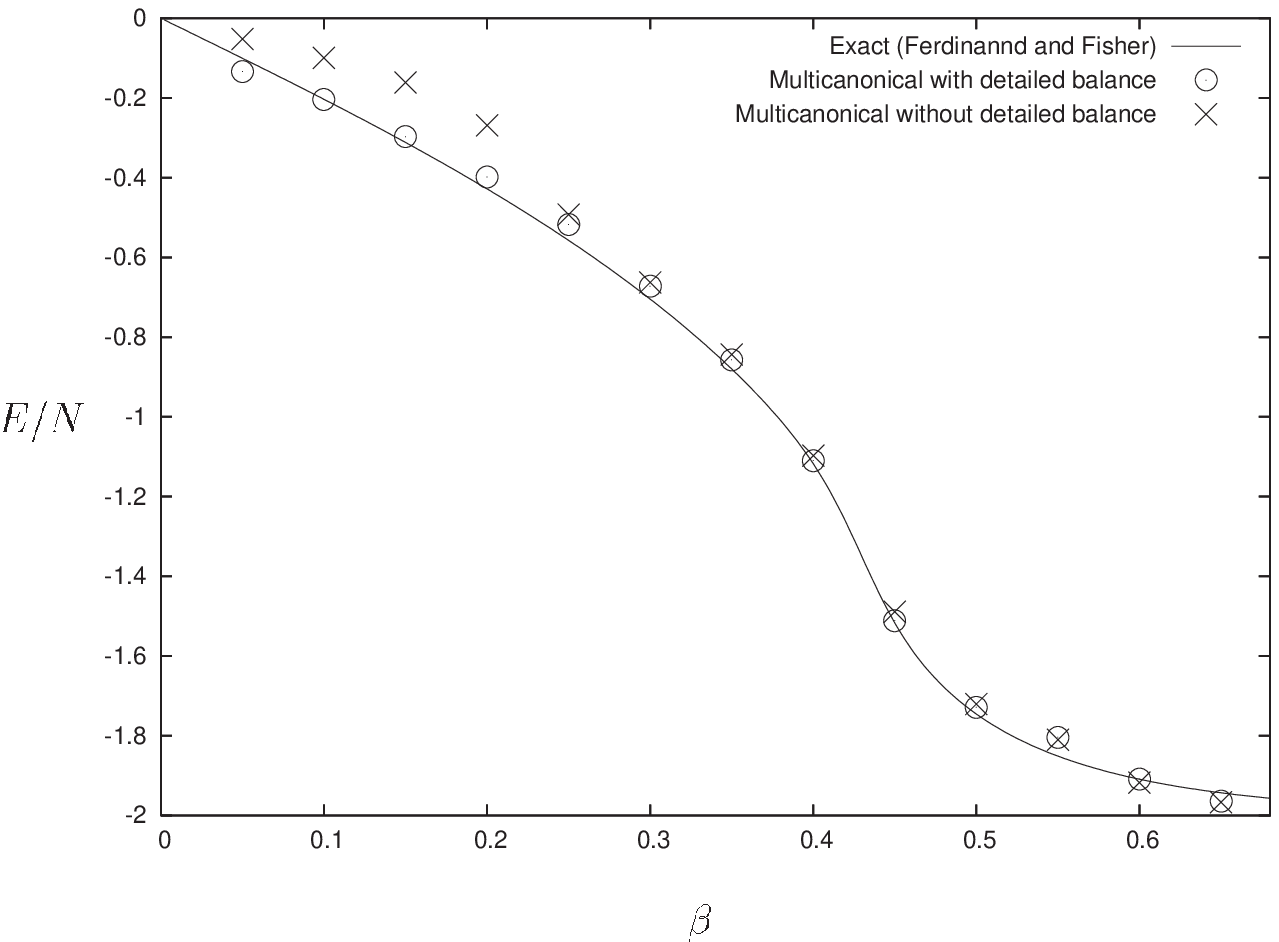}
\end{center}
\caption{Comparison of energy among the exact (solid line), 
multicanonical with the detailed balance (circles), 
and multicanonical without the detailed balance (crosses), 
where we have set $L=20$, $p=32$, $M_{\rm total}=10^8$, 
and $M_{\rm node}=100$. 
\label{energy20}}
\vspace{0.5cm}
\end{figure}

Figure \ref{energy20} is a comparison of energy 
among three results with $L=20$: exact (solid line), 
multicanonical with the detailed balance (circles), 
and multicanonical without the detailed balance (crosses). 
When the detailed balance is imposed on the joints of 
constituent Markov chains (circles), 
the energy agrees well with the exact one. 
This result is consistent with our intuition because 
it is in principle guaranteed that sampling based on 
the detailed balance gives the correct distribution. 
On the other hand, when the detailed balance is not imposed 
on the joints (crosses), there is slight deviation of energy 
from the exact one in the high-temperature region. 

According to the reweighting formula (\ref{reweighting2}), 
the low-energy part of the multicanonical weight is dominant 
when $\beta$ is large. 
In Fig. \ref{energy20}, the crosses show that the generated 
weight is sufficiently accurate in the low-energy region. 
This is because the low-energy part of the weight is generated 
in the last phase of a weight-generation process and therefore 
accumulation of errors is not large. 
On the other hand, the middle-energy part of the weight 
may accumulate errors because it is updated every iteration 
with a very short Markov chain that does not cover the 
entire energy space. 
This is the cause of the energy deviation for small $\beta$ 
when the detailed balance is not imposed on joints. 

Since even this simple Ising model produces considerable errors 
for small $M_{\rm node}$, 
one should be careful when treating more complicated systems. 
It depends on the shape of the considered multicanonical weight 
how energy deviates from the true values. 
When $M_{\rm node}$ is small due to massive parallelism, 
one can obtain better accuracy by imposing the detailed balance 
on joints between constituent Markov chains.

\begin{table}[htb]
\vspace{0.5cm}
\caption{Execution (E), operation (O), and communication (C) time 
of the code for parallelism $p=2,4,8,16,32,64$ have been measured 
with $L=100$ and $M_{\rm total}=10^8$ in the first 
iteration of weight generation. E is the sum of O and C. 
Scalability of execution time is evaluated in units of 
the $p=2$ case. 
A PC cluster with Xeon 1.50 GHz processors 
has been used for the measurements. }
\vspace{0.2cm}
\label{scalability}
\begin{tabular}{rrrrr}
\hline
$p$ & E (sec) & O (sec) & C (sec) & Scalability \\
\hline
 2 & 12.32 & 11.77 & 0.55 &  2.00 \\
 4 &  6.21 &  5.93 & 0.28 &  3.97 \\
 8 &  3.80 &  3.59 & 0.21 &  6.48 \\
16 &  2.19 &  1.66 & 0.53 & 11.25 \\
32 &  1.40 &  0.77 & 0.63 & 17.60 \\
64 &  0.97 &  0.43 & 0.54 & 25.40 \\
\hline
\end{tabular}
\vspace{1.0cm}
\end{table}

Finally, in Table \ref{scalability}, we show scalability of the code 
in the case of $L=100$ and $M_{\rm total}=10^8$, which show 
$p$-dependence of execution (E), operation (O), 
and communication (C) time. We have represented the scalability 
concerning to execution time in units of the $p=2$ case. 
The measurements have been 
performed in the first iteration of weight generation. 
As $p$ increases, execution time decreases well, but there is 
deviation from the ideal scalability. 
This is because the total number of operations is not large 
compared to communications and a small part of the operations 
have not been parallelized. Communication time is comparable 
with operation when $p$ is large. 
For better performance, serial operations and collective 
communications should be removed as much as possible. 
We could invent more sophisticated implementation
of the algorithm for complete scalability. We will consider 
improvement of the code in the next paper.

\section{Conclusions}
\label{conclusions}
We have proposed a parallelization algorithm 
for Markov-chain generation, which can be 
applied to any MCMC-based methods. 
We have verified the algorithm in the two-dimensional Ising model 
combined with the multicanonical method. We have confirmed 
accuracy of the obtained multicanonical weights by checking 
agreement of energy, specific heat, free energy, and entropy 
with the exact results. 
We have also shown that multicanonical weights may have
errors if the detailed balance is not used
for linking short Markov chains generated in parallel. 
One can decrease such errors if unnecessary configurations 
are discarded with the proposed algorithm when connecting 
constituent Markov chains. 
The algorithm will be useful for highly massive parallelism 
equipped with future supercomputers.

\section*{Acknowledgments}
The authors would like to thank Y.~Iba and 
N.~Kamiya for useful discussions and conversations. 
The numerical calculations were carried on the RIKEN 
Super Combined Cluster (RSCC) system. 

\appendix


\begin{thebibliography}{00}

\bibitem{metropolis}
N.~Metropolis, A.~W.~Rosenbluth, M.~N.~Rosenbluth, and A.~H.~Teller: 
J. Chem. Phys. {\bf 21} (1953) 1087.

\bibitem{protein}
{\it Protein Folding}, edited by T.~E.~Creighton 
(Freeman, New York, 1992). 

\bibitem{suzuki}
M.~Suzuki, S.~Miyashita, and A.~Kuroda: Prog. Theor. Phys. {\bf 58} (1977) 1377.

\bibitem{creutz}
M.~Creutz: Phys. Rev. Lett. {\bf 43} (1979) 553; 
ERRATUM Phys. Rev. Lett. {\bf 43} (1979) 890. 

\bibitem{berg1}
B.~A.~Berg and T.~Neuhaus: Phys. Lett. {\bf B267} (1991) 249; 
Phys. Rev. Lett. {\bf 68} (1992) 9.

\bibitem{berg2}
B.~A.~Berg: Fields Inst. Commun. {\bf 26} (2000) 1. 

\bibitem{iba}
Y.~Iba: Int. J. Mod. Phys. C {\bf 12} (2001) 623. 

\bibitem{hoe}
U. H. E. Hansmann, Y. Okamoto, and F. Eisenmenger: 
Chem. Phys. Lett. {\bf 259} (1996) 321.

\bibitem{nakajima}
N.~Nakajima, H.~Nakamura, and A.~Kidera: 
J. Phys. Chem. B {\bf 101} (1997) 817. 

\bibitem{ikebe}
J.~Ikebe, N.~Kamiya, H.~Shindo, H.~Nakamura, and J.~Higo: 
Chem. Phys. Lett. {\bf 443} (2007) 364.

\bibitem{kamiya}
N.~Kamiya, Y.~Yonezawa, H.~Nakamura, and J.~Higo: 
Proteins {\bf 70} (2008) 41.

\bibitem{note}
Some works have used the multiple-reweighting method 
proposed by Ferrenberg and Swendsen \cite{fs}
to make density of states in protein-folding simulations.\cite{mso}
It would be interesting to consider relationship 
between the multiple-reweighting and the current methods. 

\bibitem{fs}
A.~M.~Ferrenberg and R.~H.~Swendsen: 
Phys. Rev. Lett. {\bf 63} (1989) 1195.

\bibitem{mso}
A. Mitsutake, Y. Sugita, and Y. Okamoto: 
J. Chem. Phys. {\bf 118} (2003) 6676.

\bibitem{mpi}
http://www.mpi-forum.org/

\bibitem{ferdinand}
A.~E.~Ferdinand and M.~E.~Fisher: Phys. Rev. {\bf 185} (1969) 832. 

\end{thebibliography}
\end{document}